\documentstyle[epsf,aps,preprint,floats,tighten]{revtex} 

\def\Slash #1{\hbox{$#1$}\mkern-11.0mu\lower 0.pt \hbox{/}} 
\def\Lc #1{\cal L} 
 
\begin{document}
\begin{titlepage}
\samepage{
\rightline{IFIC/98-78}
\rightline{FTUV/98-79}
\rightline{\tt hep-ph/9810371}
\rightline{October 1998}
\vfill
\begin{center}
   {\Large \bf Basis independent study of Supersymmetry without R--parity\\ 
     \smallskip
      and the tau neutrino mass}
\vfill
\vspace{.08in}
   {\large
       Javier Ferrandis 
    \\}
\vspace{.18in}
 {\it   Institut de F\'\i sica Corpuscular - C.S.I.C.\\
Departament de F\'\i sica Te\`orica, Universitat de Val\`encia\\
46100 Burjassot, Val\`encia, Spain \\
 email: ferrandis@bigfoot.com
}
\end{center}
\vfill
\begin{abstract}
  {\rm In the most general R--parity violating model, including
both bilinear and trilinear terms, the sneutrino receives a vacuum expectation
value, $<\widetilde{\nu}>$.
We investigate the constraints on $<\widetilde{\nu}>$ 
following a basis independent approach, highlighting the relations
between the three most popular basis. 
We study the prediction for the 
tau neutrino mass which follows 
from the minimization of the scalar potential 
in a SUGRA model with universality of 
the soft parameters at the GUT scale. 
Finally we show that the tau 
neutrino mass controls the R-parity violating effects both in the fermionic 
and scalar sectors.}
\end{abstract}
\vspace{.13in}  }
\end{titlepage}
\setcounter{footnote}{0}

%%%%%%%%%%%%%%%%%%%%%%%%%%%%%%%%%%%% 
 
\newpage 
%%%%%%%%%%%%%%%%%%%%%%%%%%%%%%%%%%%% 

\section{Introduction} 
 
%%%%%%%%%%%%%% INTODUCCION %%%%%%%%% 
It is well known that if the conservation of the discrete symmetry known as 
R--parity is imposed the presence of terms in the superpotential which 
violate leptonic and baryonic number is not allowed. The existing 
phenomenological studies of explicit R--parity breaking 
have so far concentrated on 
exploring the effects of the trilinear and the bilinear terms separately. 
In this paper we study both together following a basis independent approach 
to parametrize the R--parity violating effects 
\cite{Davidson:1997cc,Davidson:1997mc,Davidson:1998yy}. It has 
been pointed out that it is not possible to eliminate completely the effects 
of the bilinear terms by simply redefining fields
\cite{Aulakh:1982yn,Hall:1984id,Lee:1984tn,Nardi:1997iy}, and so any 
complete study should include all possible R--parity violating terms. 
The studies which have concentrated solely on the effects 
of the trilinear terms \cite{Dreiner:1997uz} 
have claimed the elimination of the bilinear terms 
considering them proportional to the neutrino masses; 
in contrast one can find published papers which point out 
\cite{Nilles:1997ij,Hempfling:1996wj,Nowakowski:1996dx,Joshipura:1994ib,Diaz:1997xc,deCampos:1995av,Roy:1997bu} 
that the effects of the bilinear terms cannot 
generally be considered negligible.
One of the sources of this apparent contradiction is that we can find 
papers which quote strong bounds on the sneutrino vevs and other 
papers which tell us that these vevs can be very large. 
Hence there is some confusion in the literature 
with regard to limits on sneutrino vevs. 
In this work we show, in the 
one generation model, that the best way to understand the R--parity 
violating effects lies in following an approach which defines a
set of basis independent variables
which clearly parametrize R--parity violating effects.
Therefore any limit obtained on these variables holds in any basis.
We shall see that sneutrino vevs are not suitable to parametrize 
R--parity violation unless we clearly specify the basis in which we 
are working. Following this approach too we show, 
motivated by the recent experimental results \cite 
{Fukuda:1998mi,Fukuda:1998fd} and phenomenological studies \cite 
{Kong:1997di,Mukhopadhyaya:1998xj,Bednyakov:1998cx,Chun:1998gp,Faessler:1998db,Joshipura:1998fn,Joshipura:1997yb} 
on neutrino masses, that the $\nu _\tau $ mass controls R--parity 
violating effects both in the fermionic sector and in the scalar sector
\footnote{Some correlations between neutrino masses and lepton
number violating processes have been studied in previous works 
\cite{Nowakowski:1996dx}}
. To 
get this result we need to take into account the minimization of the scalar 
potential which allows us to connect the parameters which control the 
R-parity violating effects in the scalar sector with the parameters which 
control the effects in the fermionic sector. 
 
\section{One generation $\Slash{R}$--MSSM model} 
 
The one generation model is a useful toy model which has been used 
on various occasions   
\cite{Nilles:1997ij,Hempfling:1996wj,Nowakowski:1996dx,Joshipura:1994ib,Diaz:1997xc,deCampos:1995av,Roy:1997bu,Akeroyd:1997iq,Bisset:1998bt} 
to explore R--parity violating effects since it contains the main 
ingredients of the complete model with three generations. 
The superpotential, $W$, of the model appears as:
$$W=W_{MSSM}+W_{\Slash{R}} \ ,$$
where $W_{MSSM}$ is the familiar superpotential of the MSSM. Note
that we denote $\widehat{L}_{0}=\widehat{H}_d$, with $v_0=v_d$ and
we keep the $h_{b}$ for the case where $\lambda_0^D \propto M_{b}$:  
$$ 
W_{MSSM} = \left[ h_t\widehat{Q}_3  \widehat{H}_u\widehat{U}% 
_3+\lambda _{0}^D\widehat{L}_0  \widehat{Q}_3\widehat{D}_3 
+h_{\tau} \widehat{L}_0 \widehat{L}_3\widehat{R}_3 
-\mu_{0}\widehat{L}_0 \widehat{H}_u
\right]  
$$ 
$W_{\Slash{R}}$ breaks R--parity and is given by:
\begin{equation} 
W_{\Slash{R}}=-\mu_{3}\widehat{L}_3 \widehat{H}_u + \lambda 
_{3}^D\widehat{L}_3 \widehat{Q}_3\widehat{D}_3  
\end{equation}
The scalar potential contains some relevant terms for the 
following discussion:  
\begin{equation}
V_{soft}= \left(  
\begin{array}{c} 
\widetilde{L}_0 \\ \widetilde{L}_3  
\end{array} 
\right)^{\dagger} \left(  
\begin{array}{cc} 
M_{L_{0}}^2 & M_{L_{03}}^2 \\  
M_{L_{30}}^2 & M_{L_{3}}^2  
\end{array} 
\right) \left(  
\begin{array}{c} 
\widetilde{L}_0 \\ \widetilde{L}_3  
\end{array} 
\right) - \left( E_{\alpha} \widetilde{L}_{\alpha} H_u \right.  
\end{equation}
$$ 
\left. - A^D_{\alpha\beta } \lambda _{\beta}^D \widetilde{L}_{\alpha}  
\widetilde{Q}_3 \widetilde{D}_3 +h.c. \right)  
$$ 
We assume that $E_{\alpha}= B_{\alpha\beta} \mu_\alpha$. 
Then $B_{\alpha\beta}$ and $A^D_{\alpha\beta }$ will have covariant
properties of transformation under the $SU(2)$ symmetry of the 
$L_{\alpha}$ fields ($SU(4)$ symmetry in the three generation model).
 The three minimization equations are \footnote{% 
We assume there is no charge or color breaking minima \cite 
{Casas:1997ze,Abel:1998ie}}:  
$$ 
\left(m_u^2 - g_Z^2v_Y^2 \right) v_u- E_0 v_0- E_3v_3=0  
$$ 
\begin{equation} 
\left( m^2_{00} + g_Z^2v_Y^2 \right)v_0-v_u E_0+m_{03}^2v_3=0  
\end{equation}
$$ 
\left( m^2_{33} + g_Z^2v_Y^2 \right)v_3-v_u E_3+m_{30}^2v_0=0  
 \ ,$$ 
where:  
$$ 
m_u^2= \left( M_{H_u}^2+ \sum_{\alpha =0}^3\left| \mu \right| _\alpha ^2 
\right)  
$$ 
\begin{equation} 
m_ {\alpha \beta}^2 = \left( M_{L_{\alpha \beta}}^2 + \mu _\alpha 
\mu_{\beta}^{*} \right)  
\end{equation}
$$ 
v^2_Y=((v_0^2+v_3^2)-v_u^2)  
$$ 
Under what conditions can we rotate the $\mu _i-$terms 
without affecting the scalar potential ? 
The following two conditions are sufficient 
\cite{Nardi:1997iy,Nowakowski:1996dx,Borzumati:1996hd,Banks:1995by}: 
\begin{itemize} 
\item  $\mu _\alpha $ is an eigenvector of ${\bf {M}_L^2\,}$. 
 
\item  $E_\alpha $ is proportional to $\mu _\alpha $: $E_\alpha =B\mu 
_\alpha $. 
\end{itemize} 
These conditions are not satisfied in general at the electroweak scale,
although it may be possible to satisfy them at some unification scale 
where soft breaking parameters are universal. 
Therefore it is not possible in general to eliminate the effects of 
the bilinear terms and the main consequence will be the 
appearance of the tau neutrino mass at tree level. 

\section{Basis for R--parity. Invariants} 
 
It is useful to work in other basis to understand better 
the magnitude of the R--parity violating phenomenological effects. 
\cite{Nardi:1997iy,Bisset:1998bt,Diaz:1998vf,Diaz:1998vq} . 
We define the following rotation of the Higgs down and leptonic 
fields, $L_{\alpha}$, which in turn induces a rotation of the $\mu $% 
--terms:
\begin{equation}
\cal{O} =\left(  
\begin{array}{cc} 
s_\theta  & c_\theta  \\  
-c_\theta  & s_\theta  
\end{array} 
\right)
\label{eq:mu-vev-rot}
\end{equation}
Under this change of basis in the Lagrangian we find that the 
bilinear soft parameters and the trilinear couplings 
are rotated in the same way, in matricial form:
\begin{equation}
\begin{array}{c} 
E^{\prime} = \cal{O} E \\
B^{\prime} = \cal{O} B \cal{O}^{t} \\
(A^{D})^{\prime} = \cal{O} A^{D} \cal{O}^{t} 
\end{array}
\end{equation}
The $2\times 2$ mass matrix for the soft masses changes to 
$\cal{O}^{t} M_L^2 \cal{O}$:  
\begin{equation}
(M_L^2)^{\prime }=\left(  
\begin{array}{cc} 
s_\theta ^2M_{L_0}^2+c_\theta ^2M_{L_3}^2+s_\theta c_\theta 
(M_{L_{03}}^2+M_{L_{30}}^2) & M_{L_{03}}^2s_\theta ^2+M_{L_{30}}^2c_\theta 
^2+s_\theta c_\theta \Delta M^2 \\  
M_{L_{30}}^2s_\theta ^2+M_{L_{03}}^2c_\theta ^2+s_\theta c_\theta \Delta M^2 
& s_\theta ^2M_{L_3}^2+c_\theta ^2M_{L_0}^2+s_\theta c_\theta 
(M_{L_{03}}^2-M_{L_{30}}^2) 
\end{array} 
\right)   
\label{eq:softmass}
\end{equation}
Where $\Delta M^2=M_{L_3}^2-M_{L_0}^2$.  
From the phenomenological point of view it will be convenient to 
define the non covariant parameters $B_0$ and $B_3$ as $E_0=B_0\mu_0$
and $E_3=B_3\mu_3$ because as we will see 
the combinations $\Delta B=B_3-B_0$ and $\Delta M^2$ appear frequently
and their RGE's are simpler to analyze.
Now it is obvious that: $ E_\alpha ^{\prime }\neq \mu _\alpha 
^{\prime }B_\alpha ^{\prime }$. If we write 
$E_\alpha^{\prime}$ as a function of $B_{\alpha}$ we obtain two useful
equations:  
$$ 
E_0^{\prime }=
\left(s_{\theta}\mu_0 +c_{\theta}\mu_3\right) \left( s_\theta 
^2B_0+c_\theta ^2B_3\right) 
+s_\theta c_\theta 
\left(s_{\theta}\mu_3-c_{\theta}\mu_0\right)\Delta B  
$$ 
\begin{equation}
 E_3^{\prime }=
\left(s_{\theta}\mu_3-c_{\theta}\mu_0\right) 
\left( s_\theta^2B_3+c_\theta ^2B_0\right) +
s_\theta c_\theta 
\left(s_{\theta}\mu_0 +c_{\theta}\mu_3\right) \Delta B  
\end{equation}
The most useful basis are those which involve some simplification of 
the phenomenological studies. Assuming that our initial basis is
arbitrary, we give below the rotation angle to 
three basis of particular interest: 
\begin{itemize} 
\item  I) The basis where the trilinear coupling is zero:  
\begin{equation}
s_\theta =\frac{\lambda _0^D}{\sqrt{(\lambda _0^D)^2+(\lambda _3^D)^2}}% 
\qquad c_\theta =\frac{\lambda _3^D}{\sqrt{(\lambda _0^D)^2+(\lambda _3^D)^2}% 
}  
\label{eq:rot-basisI}
\end{equation}
\item  II) The basis where $\mu _3^{\prime }=0$:  
\begin{equation}
s_\theta =\frac{\mu _0}{\sqrt{\mu _0^2+\mu _3^2}}\qquad c_\theta =\frac{\mu 
_3}{\sqrt{\mu _0^2+\mu _3^2}}  
\label{eq:rot-basisII}
\end{equation}
\item  III) The basis where $v_3^{\prime }=0$:  
\begin{equation}
s_\theta =\frac{v_0}{\sqrt{v_0^2+v_3^2}}\qquad c_\theta =\frac{v_3}{\sqrt{% 
v_0^2+v_3^2}}  
\label{eq:rot-basisIII}
\end{equation}
\end{itemize} 
We notice that it is impossible to get at the same time 
$\mu _3^{\prime}=0$ and $v_3^{\prime }=0$.
 Basis I) is favoured if we wish to use the 
renormalization group equations. From the phenomenological point of view 
basis III) is convenient because in this basis the Higgs down, 
$\widetilde{L}_0$, is the field 
which gives mass to the bottom quark; in the general case this is not 
true since we have an additional contribution from a non-zero 
$v_3^{\prime}$ . We notice that it is easy to 
find basis independent parameters, for instance:  
\begin{equation}
v_d=\sqrt{v_0^2+v_3^2}\Rightarrow v=\sqrt{v_u^2+v_d^2}\qquad \mu =\sqrt{\mu 
_0^2+\mu _3^2}  
\end{equation}
$$ 
\lambda ^D=\sqrt{(\lambda _0^D)^2+(\lambda _3^D)^2}  
$$ 
From the above we can deduce that the generalization of the MSSM
definition of $\tan \beta$, given by
\begin{equation}
\tan \beta =\frac{v_u}{v_d} \ ,  
\end{equation}
is also a basis invariant.
There are other invariants which turn out to be very useful, 
and are defined as  
%%%%%%%%%%%%% 
\footnote{% 
We follow the same notation as ref. \cite{Nilles:1997ij} 
in defining $\sin\zeta$ 
and $\sin\gamma$.}: %%%%%%%%%%%%% 
\begin{equation} 
\label{eq:zeta}\cos \zeta =\frac{\mu _\alpha v_\alpha }{\mu v_d}\Rightarrow 
\sin \zeta =\frac{(\mu _0v_3-\mu _3v_0)}{\mu v_d} 
\end{equation} 
\begin{equation} 
\label{eq:gamma}\cos \gamma =\frac{\lambda _\alpha ^D\mu _\alpha }{\lambda 
^D\mu }\Rightarrow \sin \gamma =\frac{(\lambda _0^D\mu _3-\lambda _3^D\mu _0)% 
}{\lambda ^D\mu }=\left( \frac{\mu _3}\mu \right) ^I 
\end{equation} 
\begin{equation} 
\label{eq:chi}\cos \chi =\frac{\lambda _\alpha ^Dv_\alpha }{\lambda ^Dv_d}% 
\Rightarrow \sin \chi =\frac{(\lambda _0^Dv_3-\lambda _3^Dv_0)}{\lambda ^Dv_d% 
}=\left( \frac{v_3}{v_d}\right) ^I 
\end{equation} 
Here the index I) indicates that the expression has to be evaluated in the 
basis I) where $\lambda _3^D=0$. In ref. \cite{Davidson:1998yy} it was 
shown that some invariants are useful in obtaining cosmological bounds on 
R--parity violating effects. 
We will show that the R--parity violating invariant variables $\sin \zeta $ 
and $\sin \gamma $ are very useful to 
study the $\nu _\tau $ mass and the R--parity violating effects in general
in the fermionic sector, while $\sin \chi $ is appropiate for studying
for instance the phenomenology of the process $Z\Rightarrow b\overline{b}$. 
We will extract new experimental bounds on the 
$\widetilde{\nu}_\tau$ vev from the latter and other  
processes. In this model the masses of the vector bosons and the quark 
masses are given by:  
\begin{equation} 
\label{eq:bosonmas}M_W^2=\frac{g^2}4(v_u^2+v_d^2)=\frac{g^2}4v^2 
\end{equation} 
\begin{equation} 
\label{eq:topmas}M_t=\frac{h_t}{\sqrt{2}}v_u=\frac{h_t}{\sqrt{2}}\frac{2M_W}g% 
\frac{t_\beta }{\sqrt{1+t_\beta ^2}} 
\end{equation} 
\begin{equation} 
\label{eq:botmas}M_b=\frac 1{\sqrt{2}}\left( \lambda _0^Dv_0+\lambda 
_3^Dv_3\right) =\frac{c_\chi \lambda ^D}{\sqrt{2}}\frac{2M_W}g\frac 1{\sqrt{% 
1+t_\beta ^2}} 
\end{equation} 
We can see that the expressions on the right hand side are basis independent 
because the masses are physical observables. We keep the expression $h_b$ 
for the case where $\lambda _0^D\propto M_b$.\\
{\bf Basis in the three generation model}.
Basis II) and III) can be straightforwardly generalized to
more generations. An arbitrary basis 
is not useful unless we can define it unambiguously.
We can see that in the three generation model
the only useful generalization of basis I) 
is the basis linked through the renormalization
group equations to the high energy basis 
(SUGRA basis with universality) where the $\mu_i$ 
terms have been rotated away. This basis can 
be useful for finding relations with the SUGRA parameter
space.We must point out that in a three generation 
model there is no basis which simplifies the RGE's 
as in the one generation model.

\section{One generation SUGRA $\Slash{R}$--MSSM} 
 
{\bf $\lambda _3^D$--approach.} The one generation $\Slash{R}$--MSSM 
model is completely fixed if we rotate at the GUT scale 
to some basis where, if there is universality of the soft parameters, 
we eliminate the bilinear terms completely both in the 
superpotential and in the soft sector. 
In this basis all R--parity violating 
effects are fixed if $\lambda _3^D(GUT)$ is known. 
Then we use the RGE's to compute the parameters at the 
electroweak scale. We can see that the presence of 
trilinear parameters induces the appearance 
through RGE's evolution of a vev 
because at the electroweak scale 
the parameters $\mu _3$,$M_{L_{03}}^2$ and $M_{L_{30}}^2$ will
be non--zero.  
\begin{equation}
\begin{array}{ccccc} 
\begin{array}{c} 
\mu _3(GUT)=0 \\  
\lambda _3^D(GUT)\neq 0 
\end{array} 
& \stackrel{RGE}{\Longrightarrow } & \left\{  
\begin{array}{c} 
\mu _3\neq 0 \\  
E_3 \neq 0 \\  
M_{L_{03}}^2\neq 0 \\  
M_{L_{30}}^2\neq 0 \\  
\lambda _3^D\neq 0 
\end{array} 
\right.  & \Rightarrow  & v_3\neq 0 
\end{array} 
\end{equation}
{\bf $\mu _3$--approach.} Another possibility is to rotate 
the fields at the GUT\ scale to a basis 
where $\lambda _3^D(GUT)=0$, and make use of the fact that 
the RGE's satisfy the following properties: 
\begin{equation}
\frac{d\lambda _3^D}{dt}\propto \lambda _3^D\qquad \frac{dM_{L_{03}}^2}{dt}% 
\propto (M_{L_{03}}^2,M_{L_{30}}^2,\lambda _3^D)  
\end{equation}
There is also an analogous equation for $M_{L_{30}}^2$. 
The Lagrangian parameters at the electroweak scale can then 
be computed, and one finds:  
\begin{equation}
\begin{array}{ccccc} 
\begin{array}{c} 
\mu _3(GUT)\neq 0 \\  
\lambda _3^D(GUT)=0 
\end{array} 
& \stackrel{RGE}{\Longrightarrow } & \left\{  
\begin{array}{c} 
\mu _3\neq 0 \\  
E_3 \neq 0 \\  
M_{L_{03}}^2=0 \\  
M_{L_{30}}^2=0 \\  
\lambda _3^D=0 
\end{array} 
\right.  & \Rightarrow  & v_3\neq 0 
\end{array} 
\end{equation}
Following this approach we get at the electroweak scale the parameters in 
the basis I). As we can see all the R-parity violating effects are completely 
fixed by $\mu _3(GUT)$. It does not matter which approach we are following 
because both are equivalent. The main point is that if we do not neglect 
any contributions in the superpotential we can rotate from 
one basis to another 
and thus compare calculations which have been 
performed in different basis. These results 
have to be the same. To build a SUGRA extension of the one generation 
$\Slash{R}$--MSSM model we need to look for basis independent parameters. In 
any case the most correct parameters are:
$$ 
\left( A_0,M_0,M_{1/2},t_\beta ,\mu _3^G\right)   
$$ 
in the basis I)\ where $\lambda _3^D(GUT)=0$. Or:  
$$ 
\left( A_0,M_0,M_{1/2},t_\beta ,(\lambda _3^D)^G\right)   
$$ 
in the basis II) where $\mu _3(GUT)=0$. We notice that the soft masses $% 
M_{L_{03}}^2$ and $M_{L_{30}}^2$ are zero in any basis at the GUT\ scale if 
there is universality. In the MSSM\ we need in addition to the first four 
parameters the sign of $\mu _0$ because the CP--odd scalar 
mass satisfies at tree 
level the following relation $M_A^2=t_\beta \mu _0B_0$ . Then the sign of $% 
B_0$ is fixed and we can compute it through the minimization equations. 
In this model, as we will see later, the relation between 
the CP--odd mass and the minimization
equations is more complex because there are additional 
contributions and there is mixing between neutral Higgses and sneutrinos. 
Now we need to specify $\mu _3^G$ including its sign, and this will allow us 
to compute $\mu _0$, $B_0$ and $B_3$ through 
the minimization equations. 
We have complemented the analysis of this work with a 
numerical study following the $\mu _3$--approach 
which simplifies the RGE's substantially but we have expressed
our results in terms of basis independent parameters.\\ 
{\bf Notation.} From now on we will denote the parameters in 
basis I) as $\mu _3$, $v_3$ and $\lambda _\alpha ^D$ 
and the parameters in the other basis using an index II) 
or III). In addition we will use $v_3^{\prime }$ to denote the vev in the 
basis II) and $\mu _3^{\prime }$ to denote the $\mu _3$--term in the basis 
III) to avoid confusion.
%%%%%%%%%%%%%%%%%%%%%%%%%%% 
\begin{table} \centering    
\begin{tabular}{|c|c|c|} \hline 
  I                &    II                 &      III            \\ \hline     $\mu_3     \neq 0$ & $\mu_3        =0$     & $\mu_3        \neq 0$ \\ 
$E_3       \neq 0$ & $ E_3         \neq 0$ & $E_3          \neq 0$ \\ 
$M_{L_{03}}^2 = 0$ & $M_{L_{03}}^2 \neq 0$ & $M_{L_{03}}^2 \neq 0$ \\ 
$M_{L_{30}}^2 = 0$ & $M_{L_{30}}^2 \neq 0$ & $M_{L_{30}}^2 \neq 0$ \\ 
$\lambda^D_3  = 0$ & $\lambda^D_3  \neq 0$ & $\lambda^D_3  \neq 0$ \\ 
$v_3       \neq 0$ & $v_3          \neq 0$ & $v_3              =0$ \\ \hline  
\end{tabular} 
  \caption{ Most useful basis for R--parity at the electroweak scale}    
  \label{Rpbasis}    
\end{table} 
%%%%%%%%%%%%%%%%%%%%%%%%%%%%%%%%%%% 
 
\section{Constraints on sneutrino vevs} 
 
%%%%%%%% DISCUSION COTAS EN DIFERENTES BASES %%%%%%%%% 
%%%%%%%%%%%%%%%%%%%%%%%%%% 
\begin{figure}[ht] 
\centerline{ \epsfxsize 5.8 truein \epsfbox {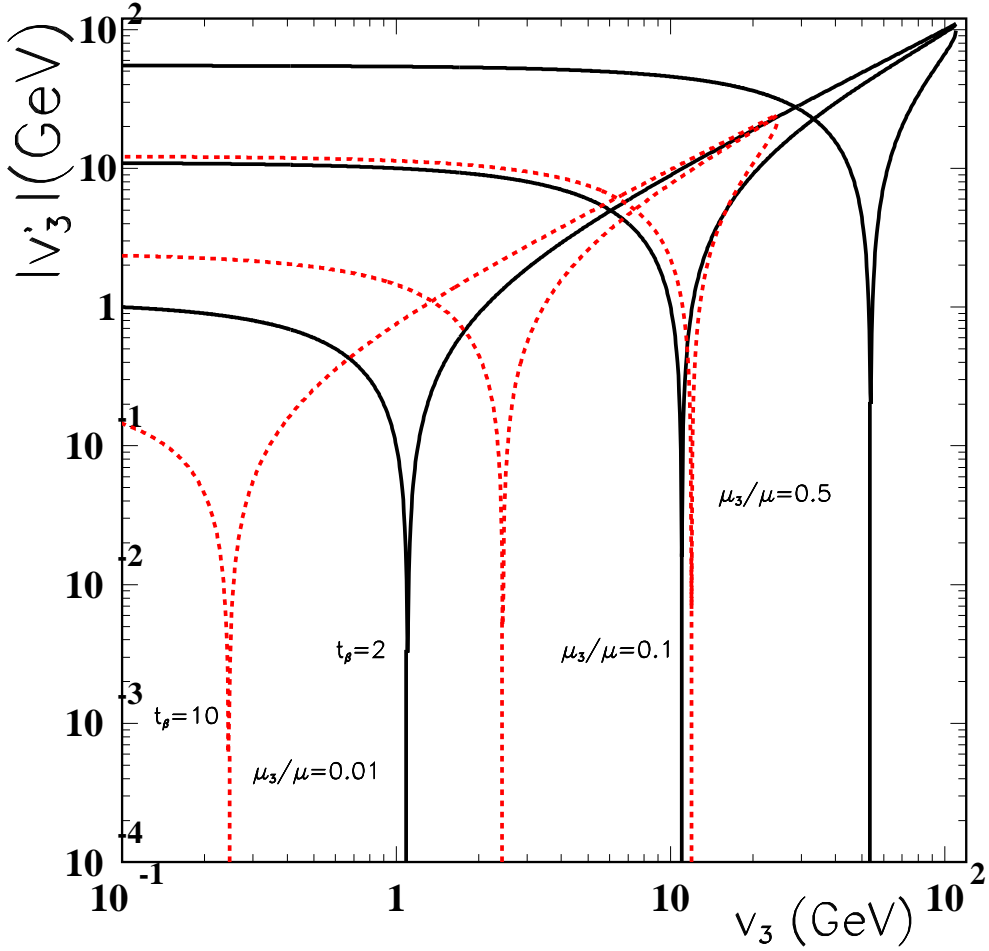}} 
\caption{\it 
Relation between the $\widetilde{\nu}_{\tau}$ vev in
different R--parity violating basis}  
\label{fig:cotas}  
\end{figure}  
%%%%%%%%%%%%%%%%%%%%%%%%%% 
Suppose we have obtained a bound on 
$\left\langle \widetilde{\nu }_\tau \right\rangle $ in the basis I) 
( where $\lambda _3^D=0$ ). 
If we wish to translate this bound to 
the basis II) we need to use the equations 
(\ref{eq:mu-vev-rot}). Then we obtain:  
\begin{equation} 
v_3^{\prime }=v_3\left( 1-\left( \frac{\mu _3}\mu \right) ^2\right) ^{1/2}-% 
\frac{\mu _3}\mu \left( \frac{v^2}{(1+t_\beta )}-v_3^2\right) ^{1/2}  
\end{equation}
We notice that to fix $v_3^{\prime }$ we need only $t_\beta $ and $\mu 
_3/\mu $ 
\footnote{We recall that $(\mu_3/\mu)^I=s_{\gamma}$ 
is an invariant.}. 
In figure (\ref{fig:cotas}) we can see the relation between
$v_3$ an $v_3^{\prime}$ for two different values of $t_{\beta}$ 
($t_{\beta}=2$ in continous line and $t_{\beta}=10$ in dashed line)
and three different values of the ratio $\mu_3/\mu$ for each value
of $t_{\beta}$.
As we can observe in figure (\ref{fig:cotas}) 
even if $v_3^{\prime }$ is small the vev in the 
basis I) can be very large depending on the values of 
$\mu _3/\mu $ and $t_\beta $. 
Then we can see that as $v_3^{\prime }=vs_\beta s_\zeta $, the smaller
the value of $v_3^{\prime }$ the smaller $s_\zeta $ will be 
and more fixed is $v_3$ , 
or in other words, greater is the adjustment 
between $\mu _3/\mu $ and $v_3/v_d$ to 
get a small value for $s_\zeta $. 
When we study the $\nu _\tau $ mass 
we will see to what extent this adjustment is possible.\\  
\begin{figure}[ht] 
\centerline{ \epsfxsize 5.8 truein \epsfbox {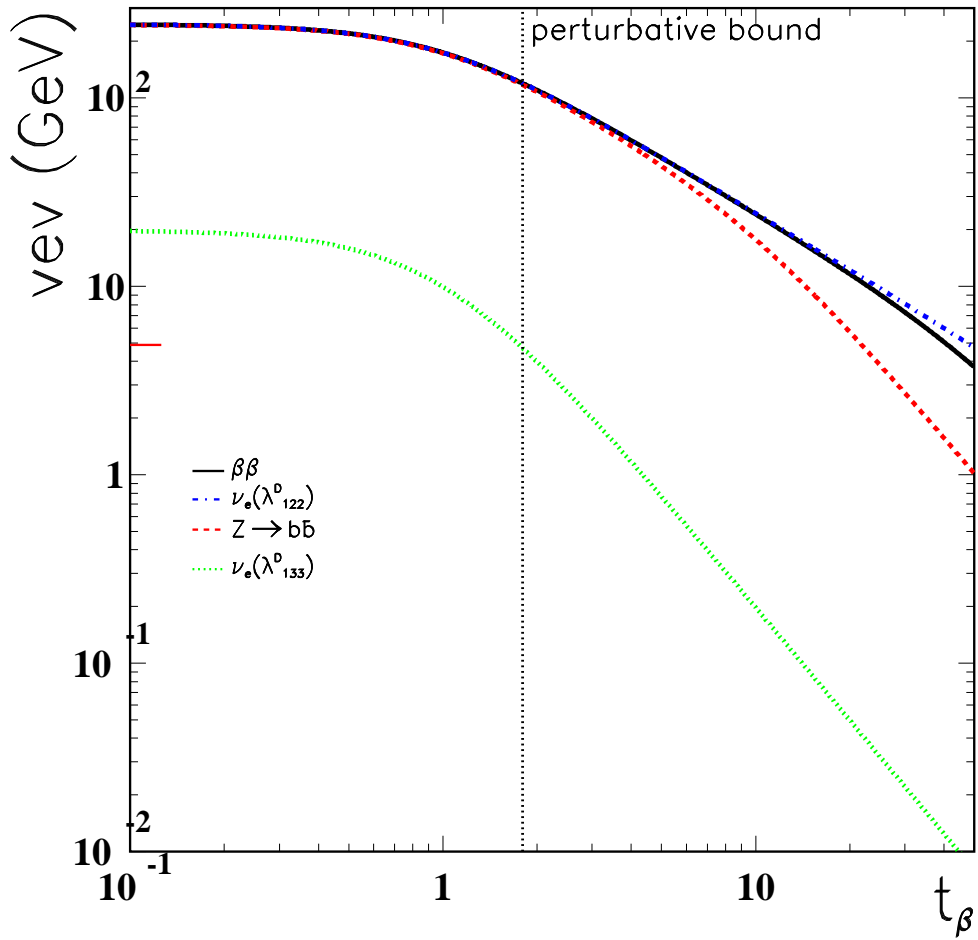}} 
\caption{\it Constraints on $\widetilde{\nu}$ vev from 
$Z\Rightarrow b\overline{b}$, double beta decay and the bound on
$\nu_e$ mass.}  
\label{fig:lambda}   
\end{figure}  
%%%%%%%%%%%%%%%%%%%%%%%%%% 
The trilinear couplings $(\lambda _0^D,\lambda _3^D)$ transform 
in the same way as the vevs and this leads us 
to the question of the experimental bounds on 
the coupling $\lambda _3^D$ because we could generate after  
a rotation a trilinear coupling that is experimentally constrained. 
In this way we could derive experimental bounds on the 
sneutrino vevs and lepton number violating  
$\mu -$terms from the known bounds on trilinear couplings. 
It is possible to derive an experimental bound 
on $\lambda _3^D$ from the LEP\ measurements of 
the process ${Z}\rightarrow b\overline{b}$ 
\cite{Bhattacharyya:1995pr,Bhattacharyya:1995bw} 
where the trilinear coupling $L_3Q_3D_3$ 
contributes radiatively. This bound has been derived in the limit 
where the radiative contributions coming from the bilinear terms
are neglected. 
The bilinear terms contribute through new loops originating from the mixing 
between charged higgses and sleptons \cite{Roy:1997bu,Akeroyd:1997iq}. The 
results of the refs \cite{Bhattacharyya:1995pr,Bhattacharyya:1995bw} 
concluded that one can derive the bound $\lambda _3^D\le 0.26(0.45)$ 
at 1(2) $\sigma$ respectively (if the squarks are $\approx 100$ GeV). 
The basis where the radiative contributions coming from 
the bilinear terms and the sneutrino vevs is minimal is the basis III) 
( where the vev is zero ). The relation between the sneutrino 
vev in the basis I) and the 
R-parity violating trilinear terms in the basis III) is obtained 
from the invariant $\sin \chi $:  
\begin{equation} 
\sin \chi =\left( \frac{v_3}{v_d}\right) ^I=\left( \frac{\lambda _3^D}{% 
\lambda ^D}\right) ^{III}  
\end{equation}
Since the R-parity violating contribution to the process $Z\Rightarrow b% 
\overline{b}$ is a basis independent we can compute the parameters at the 
electroweak scale from the GUT scale parameters 
following the $\mu _3$--approach, and afterwards rotate 
to the basis III) (where $v_3^{\prime }=0$). Then we can make use of 
the identity $\lambda ^D=(\lambda _0^D)^I=h_b$ to obtain:  
\begin{equation}
(\lambda _0^D)\frac{v_3}{\sqrt{v_0^2+v_3^2}}=\frac{\sqrt{2}m_b}{v_0}\frac{v_3% 
}{\sqrt{v_0^2+v_3^2}}\leq \lambda _M  
\end{equation}
Here $\lambda_M$ denotes the experimental bound on $\lambda^D_3$.
Then we get a bound on $v_3^I$:  
\begin{equation}
v_3^2\leq \frac{\lambda _M^2v^4}{\left( 2m_b^2(1+t_\beta ^2)+\lambda 
_M^2v^2\right) }\frac 1{(1+t_\beta ^2)}  
\end{equation}
From this inequality we notice that making use of the experimental bound, $% 
\lambda _M$, allows us to get  a bound 
on $v_3\,\,.$
This bound depends on $t_\beta $ with the maximum allowed value  
being achieved with the minimum allowed value for $t_\beta.$ 
The minimum value for $t_\beta \,$ comes from the 
perturbativity bound on the top quark yukawa coupling, or in other words, 
$h_t$ grows with the energy scale and the maximum value at the 
GUT scale is $\sqrt{4\pi }$ 
which is related to the minimum of $t_\beta $ ($\approx 1.8$) as we can see 
from the formula (\ref{eq:topmas}). 
In figure (\ref{fig:lambda}) we can see 
that the experimental bound on $v_3\,$ coming from the minimum value for 
$t_\beta$ is 115 GeV approximately. 
Moreover the larger $t_\beta \,$ is the stronger
the bound on $v_3$ will be. If $t_\beta =10$ 
the bound on $v_3$ is approx. 25 GeV. 
To translate this bound to basis II) we need to know the tau neutrino mass:
However since $v^{\prime}_3$ is generally less than $v_3$ any
bound on $v_3$ will be stronger for $v_3^{\prime}$.\\
{\bf Constraints on sneutrino vevs in the three generation model}.
This method can be generalized to obtain constraints on sneutrino vevs 
in the three generation model. Since we are interested in obtaining
upper bounds on the vevs we can simplify the calculation
assuming that in the basis where the vevs are maximum the 
R--parity violating trilinear couplings are zero. Then if we rotate
to basis III) (where sneutrino vevs are zero) we generate 
R--parity violating trilinears:
\begin{equation}
(\lambda^D_{1ij})^{\prime}=
-\sqrt{\frac{v^2_1+v^2_2+v^2_3}{v_d^2}}
\lambda^D_{0ij}
\end{equation}
From this we obtain a bound on sneutrino vevs assuming that
$\lambda_{0ii}^D$ is proportional to 
the mass of the down quark of the ith-generation:
\begin{equation}
(v^2_1+v^2_2+v^2_3) \leq 
\frac{\lambda_{1ii}^2v^4}{\left( 2m_{q_i}^2(1+t_\beta ^2)+
\lambda_{1ii}^2v^2\right) }\frac 1{(1+t_\beta ^2)}  
\end{equation}
 Finally we can use the
bounds on $(\lambda^D_{1ii})^{\prime}$ from double 
beta decay and electron neutrino mass \cite{Dreiner:1997uz}.
The strongest upper
bound cames from $(\lambda^D_{133})^{\prime}$ 
(contribution to the radiative electron neutrino mass):
\begin{equation}
\sqrt{v^2_1+v^2_2+v^2_3} \leq 5 \ \hbox{GeV}
\left( \frac{\widetilde{M}}{100 \hbox{GeV}} \right)^{1/2}
\end{equation}
Where $\widetilde{M}$ is some specific scalar mass.
All these constraints are plotted in figure (\ref{fig:lambda})
as a function of $t_{\beta}$.
In the general case these bounds are valid
if we assume that there are no cancellations 
between the different contributions to 
$(\lambda^D_{1ij})^{\prime}$.

%%%%%%%%%%%%%%%%%%%%%%%%%% 
 
\section{Tau neutrino mass and R-parity violating effects} 
 
To make clearer the relation between the tau neutrino mass 
and R-parity violating effects in 
the fermionic sector it is useful to rotate to the basis III) 
($v_3=0$) where R-parity violating 
effects are more evident. In this basis we have the following: 
$v_0^{\prime }=v_d,\mu _3^{\prime }=
\mu \sin \zeta ,\mu _0^{\prime }=\mu \cos \zeta $. 
Written in the basis III) the neutralino-neutrino 
Majorana mass matrix takes the form:  
$$ 
{\bf M}_N=\left[  
\begin{array}{ccccccc} 
M_1 & 0 & \frac 12g^{\prime }v_u & -\frac 12g^{\prime }v_d & 0 &  &  \\  
0 & M_2 & -\frac 12gv_u & \frac 12gv_d & 0 &  &  \\  
\frac 12g^{\prime }v_u & -\frac 12gv_u & 0 & -\mu \cos \zeta  & -\mu \sin 
\zeta  &  &  \\  
-\frac 12g^{\prime }v_d & \frac 12gv_d & -\mu \cos \zeta  & 0 & 0 &  &  \\  
0 & 0 & -\mu \sin \zeta  & 0 & 0 &  &  
\end{array} 
\right]   
$$ 
There are several contributions to the tau neutrino mass: the bilinear term 
in the superpotential $\mu _\alpha \widehat{L}_\alpha \widehat{H}_u$ which 
provides a mass for the fermionic component of one combination of the 
doublets $L_\alpha $ (the Higgsino); the vacuum expectation values, $% 
\left\langle L_\alpha \right\rangle =v_\alpha /\sqrt{2}$ , which contribute 
to the spontaneous breaking of the electroweak symmetry, induce a mixing 
between the neutral members of the fermion doublets, $\L _\alpha ,$ and the 
neutral gauginos. Therefore the neutrino gets a mass 
and under the experimental 
assumption $M_{\widetilde{\chi }_1^0}>>M_{\nu _\tau }$ we can derive an 
approximate formula for the $\nu _\tau $ mass:  
\begin{equation} 
\label{eq:neumas}M_{\nu _\tau }^{tree}=\frac{M_Z^2M_{\widetilde{\gamma }}\mu 
s_\zeta ^2c_\beta ^2}{\left( M_Z^2M_{\widetilde{\gamma }}s_{2\beta }c_\zeta 
-M_1M_2\mu \right) }
\end{equation} 
On the other hand in an arbitrary basis the trilinear term contributes 
radiatively \cite{Borzumati:1996hd,Babu:1995vh,Barbieri:1990qj,Roulet:1991wx} 
through a quark-squark loop. 
This contribution, 
$M_{\nu _\tau }^{\lambda_3^D},$ is proportional to $(\lambda_3^D)^2$. 
The bilinear terms induce a mixing between different fields which 
contribute to the same effective couplings giving new radiative 
contributions \cite{Hempfling:1996wj} which we denote as $M_{\nu _\tau }^{Sp} 
$ (We will not analyze here how these different radiative 
contributions depend on the SUGRA\ parameter space). 
To sum up, the tau neutrino mass at one loop 
depends on three components:  
$$ 
M_{\nu _\tau }=M_{\nu _\tau }^{tree}+\left( M_{\nu _\tau }^{\lambda 
_3^D}+M_{\nu _\tau }^{Sp}\right) ^{1-loop}  
$$ 
The contribution to the tau neutrino mass at each order of perturbation 
theory is a basis independent. At tree level this fact is obvious through 
the formula (\ref{eq:neumas}) which depends only on basis independent 
parameters. 
At the one loop level the basis independence is manifested 
if we consider all the 
radiative contributions, both bilinear and trilinear. If we rotate to the 
basis I) ($\lambda _3^D=0$)  the contribution coming from the trilinear
terms is transferred to the radiative contribution 
coming from the bilinear terms 
because as is obvious the tau neutrino mass is a basis independent. We can 
notice too that the MSSM\ limit is recovered when the $\nu_\tau $ mass 
tends to zero because this implies that $\sin \zeta $ and $\lambda _3^D$ 
tend to zero.  
$$ 
\begin{array}{ccccc} 
\begin{array}{c} 
M_{\nu _\tau }\rightarrow 0 
\end{array} 
& \Longrightarrow  & \left\{  
\begin{array}{c} 
\sin \zeta \rightarrow 0 \\  
\lambda _3^D\rightarrow 0 
\end{array} 
\right.  & \Rightarrow  & \hbox{MSSM}
\end{array} 
$$ 
It is also apparent in the basis III) that the R--parity violating effects in 
the charged lepton sector depend on the tau neutrino mass. The smaller the 
tau neutrino mass, the smaller will be these effects since the 
charged lepton Dirac mass matrix is:  
$$ 
{\bf M}_C=\left[  
\begin{array}{ccc} 
M_2 & \frac 1{\sqrt{2}}gv_u & 0 \\  
\frac 1{\sqrt{2}}gv_d & \mu \cos \zeta  & 0 \\  
0 & \mu \sin \zeta  & \frac 1{\sqrt{2}}h_\tau v_d 
\end{array} 
\right]   
$$ 
The tau Yukawa coupling is a basis independent 
which is related to the tau neutrino 
mass through the exact tree level formula \cite{Akeroyd:1997iq}:  
$$ 
h_\tau ^2=\frac{2M_\tau ^2}{v_d^2}\left\{ 1-s_\zeta ^2f(M_2,\mu ,t_\beta 
,c_\zeta )\right\}   
$$ 
The chargino masses are, in the limit of $s_\zeta $ tending to zero, 
approximately given by:  
\begin{equation}
M_{\widetilde{\chi }^{\pm }}^{\Slash{R}}=M_{\widetilde{\chi }^{\pm 
}}^{MSSM}-s_\zeta ^2\frac \mu 2\left\{ 1\mp \frac{(M_2-\mu )}{\sqrt{(M_2-\mu 
)^2+4M_W^2s_{2\beta }}}\right\}   
\end{equation}
As we can see the invariant magnitude $\sin \zeta $ 
is a key parameter of the 
model. Now it is clear that to  understand the reach of the bilinear
R--parity violating effects in the fermionic sector it is crucial 
to know the attainable values of $\sin \zeta$. We will study in a SUGRA 
scenario how the tau neutrino mass (through $\sin\zeta$) 
is related with the SUGRA parameter space.\\ 
 
%%%%%%%%% DISCUSION SOBRE LAS ECUACIONES DE MINIMO %%%%%%%%%%%% 

{\bf Minimization equations and universality of soft parameters}. 
It is possible to derive a formula for $\sin\zeta$ using the 
minimization equations of the scalar potential.
This formula will be useful for studying the attainable values of 
$\sin\zeta$ in a SUGRA inspired scenario. 
Moreover, $\sin\zeta$ is a basis independent,
therefore it does not matter which basis has been 
used to derive the relation between $\sin\zeta$ and 
the SUGRA parameter space. We will assume that we have
computed all the parameters at the electroweak scale in the 
basis I) following the $\mu_3$--approach to R--parity through 
the RGE's from GUT to weak scale.
Then we rotate to basis II) (where $\mu _3^{\prime }=0$) and we get
directly a formula for $\sin\zeta$ (because $\sin\zeta=v_3^{\prime}/v_d$). 
The relevant minimization equation is:
\begin{equation}
( M_{L_3}^2+g_Z^2v_Y^2 - \frac{\mu_3^2}{\mu^2}\Delta M^2) v_d \sin\zeta - 
v_u \frac{\mu_0 \mu_3}{\mu} \Delta B + 
\frac{\mu_0\mu_3}{\mu^2} \Delta M^2 v_d \cos\zeta =0  
\end{equation}
%%%%%%%%%%%%%%%%%%%%%%%%%% 
\begin{figure}[ht] 
\centerline{ \epsfxsize 5.8 truein \epsfbox {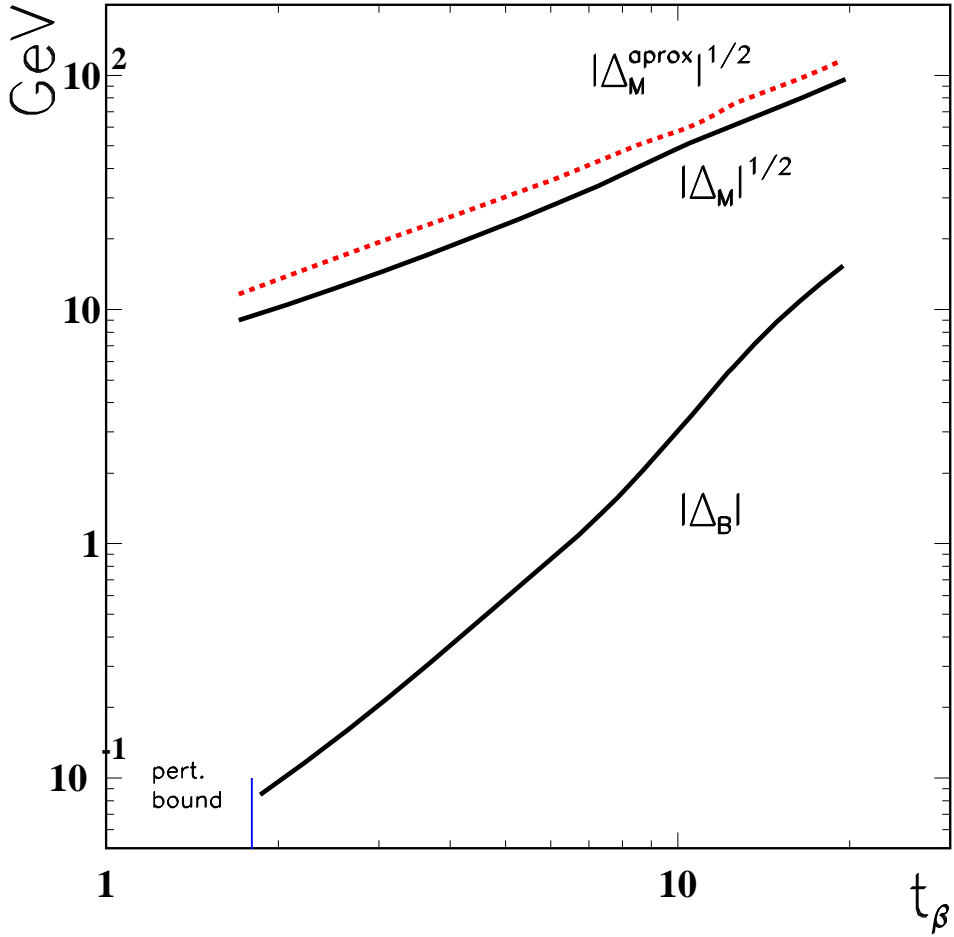}} 
\caption{\it Soft terms which contribute to the invariant $s_{\zeta}$} 
\label{fig:deltam} 
\end{figure}  
%%%%%%%%%%%%%%%%%%%%%%%%%% 
From the above one obtains a quadratic equation for $s_{\zeta}$.
We find that the invariant $s_{\zeta}$ which appeared in the fermionic
mass matrices is proportional to another invariant $s_{\gamma}=\mu_3/\mu$ 
which will be useful for studying the $\nu_{\tau}$ mass. 
At first order in $\mu_3/\mu $ the expression for $\sin \zeta$ reduces to:  
\begin{equation}
\sin\zeta= \frac{\mu _0\mu _3}{\mu^2} \left( \delta_B t_{\beta} \pm \delta_M 
\right)= \frac{1}{2}\sin(2\gamma) \left( \delta_B t_{\beta} \pm \delta_M 
\right)  
\end{equation}
where we define $M^2_{\widetilde{\nu}_3}= M_{L_3}^2+g_Z^2v_Y^2$ and:  
\begin{equation}
\delta_B = \frac{\mu \Delta B } {\left( M_{\widetilde{\nu_3 }}^2 -% 
\frac{\mu_3^2}{\mu^2}\Delta M^2 \right)} \qquad \delta_M = \frac{\Delta M^2} 
{\left( M_{\widetilde{\nu_3 }}^2 -\frac{\mu_3^2}{\mu^2}\Delta M^2 \right)} 
\ ,  
\end{equation}
We can see that
the parameters which appear on the right hand side 
$\Delta B$, $\Delta M$, $\mu_3$
and $\mu$ have to be computed in the basis I),
although the important point is that these parameters 
are fixed if we know the SUGRA parameters 
(basis independent).
There are some limiting cases: if $\delta_M \approx 0$ then $% 
\sin\zeta = \frac{\mu _0\mu _3}{\mu^2}\delta_B t_{\beta}$ or if $\delta_B 
t_B \approx 0$ then $\sin\zeta = \frac{\mu _0\mu_3}{\mu^2}\delta_M (1+  
\frac{\mu _0\mu _3}{\mu^2} \delta_M^2)^{-1/2} $ and in principle 
it would also be possible to get a very small $s_{\zeta}$ through a fine 
tuning between $\delta_Bt_{\beta}$ and $\delta_M$; 
in the latter case there could be large R--parity violating
effects in the scalar sector but small
ones in the fermionic sector. 
However we will see that this scenario is practically impossible
because $\Delta B$ is one order of magnitude less than $\Delta M$.
In order to predict the tau neutrino mass in a universality scenario
we need to study the attainable values of 
$\Delta M $ and $\Delta B$. 
Both parameters are zero at the GUT scale and they are generated radiatively.
The main question is if they could be small enough to explain a very
light neutrino.
Using the renormalization group equations (see appendix) we find
(We recall that in the $\mu_3$--approach one can take
$\lambda_3^D=0$ and $h_b=\lambda_0^D$):  
$$ 
8 \pi^2 \frac{d \Delta M^2}{d t} = - 3 h_b^2 ( M_{L_0}^2 + M_Q^2 + M_D^2 + 
A_b^2 )  
$$ 
\begin{equation}
16 \pi^2 \frac{d \Delta B}{d t} = - 6 h_b^2 A_b  
\end{equation}
We can perform an approximate analytical integration to get:
\begin{equation}
\Delta M^2(M_Z)\simeq \frac{t_U 3 h_b^2}{8\pi ^2}\left( 3M_0^2+A_b^2\right) 
\qquad \Delta B(M_Z)\simeq \left( \frac{t_U3h_b^2}{8\pi ^2}\right) A_b  
\end{equation}
Expanding the common term in the previous equations and making use of
the formula (\ref{eq:botmas}) we get:  
\begin{equation} 
\label{eq:deltam-aprox}(\Delta M^2(M_Z))^{aprox} = \frac{3g^2}{16\pi^2}\frac{% 
m^2_b}{M^2_W} ln \left(\frac{M_G}{M_Z}\right)(3M^2_0+A_b^2)\left(1+t_{\beta}^2% 
\right)  
\end{equation} 
We can see that the term $\Delta B$ is generally less than $\Delta M$:
\begin{equation}
\frac{\Delta B }{\sqrt{\left|\Delta M^2\right|}} \simeq 
\left[ 
\frac{3g^2}{16\pi^2}\frac{m^2_b}{M^2_W} 
ln \left(\frac{M_G}{M_Z}\right)\left(1+t_{\beta}^2\right)
\right]^{1/2} 
\left[1+3\frac{M_0^2}{A_b^2}\right]^{-1/2} \leq  
\frac{5}{100}\left(1+t_{\beta}^2\right)^{1/2}  
\end{equation}
Here the term $5/100$ comes from rounding off the constant term. 
The exact numerical calculations of $\Delta B$ and $\Delta M$ 
confirm this analytical estimation as can be seen in figure  
(\ref{fig:deltam}) where the numerical calculations are shown as a continous
line and the analytical approximation to $\Delta M$ is shown as a dotted
line.
We notice too that there is a minimum value for $\delta_M$ 
since there is a minimum value for $t_{\beta}$ 
which comes from the perturbativity bound on the top quark Yukawa coupling.
Then we can see that one can find the dominant relation between
the tau neutrino mass and the SUGRA space if we use the formula 
(\ref{eq:deltam-aprox}) inside the formula for $s_{\zeta}$ 
where we are neglecting the contribution coming from $\delta_Bt_{\beta}$.
Then we find:
$$ 
M_{\nu_{\tau}} \approx \delta_M^2\left(\frac{\mu_0\mu_3}{\mu^2}\right)^2  
\frac{ M_Z^2 M_{\widetilde{\gamma }} \mu c_\beta^2} {\left( M_Z^2M_{% 
\widetilde{\gamma }}s_{2\beta }-M_1M_2\mu \right)}  
$$ 
If we insert the expression (\ref{eq:deltam-aprox}) in the above 
expression we find:  
\begin{equation} 
\label{eq:neuapro2}M_{\nu_{\tau}} \approx 
\left[ 
\frac{3g^2}{16\pi^2}\frac{m^2_b}{M^2_W} 
ln \left(\frac{M_G}{M_Z}\right) \left(3 + \left(\frac{A_b}{M_0}\right)^2
\right)\right]^2 
\left(1+t_{\beta}^2\right) \left(\frac{\mu_0\mu_3}{\mu^2}\right)^2 
\frac{M_Z^2 M_{\widetilde{\gamma }} \mu} {\left( M_Z^2M_{% 
\widetilde{\gamma }}s_{2\beta }-M_1M_2\mu \right)}  
\end{equation} 
%%%%%%%%%%%%%%%%%%%%%%%%%% 
\begin{figure}[ht] 
\centerline{ \epsfxsize 5.8 truein \epsfbox {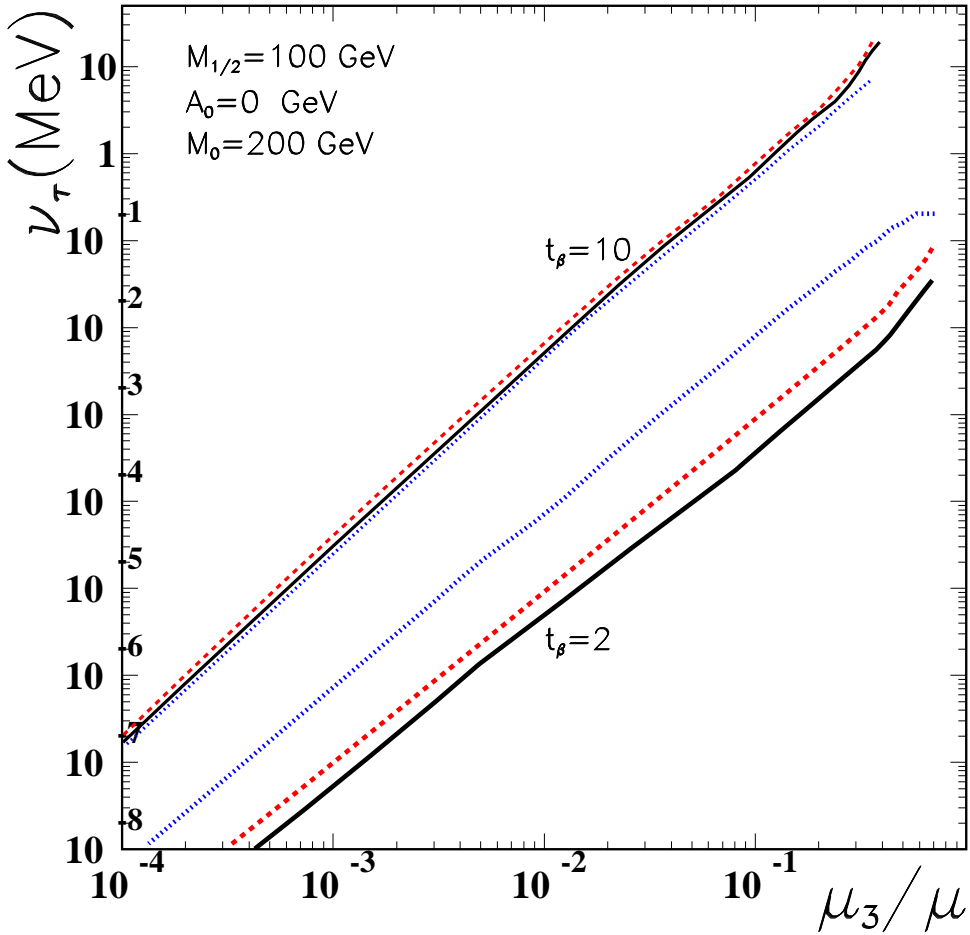}} 
\caption{\it $\nu_{\tau}$ mass as a function of the invariant 
$s_{\gamma}=(\mu_3/\mu)^I$ for two different values of $t_{\beta}$} 
\label{fig:neut-mass} 
\end{figure}  
%%%%%%%%%%%%%%%%%%%%%%%%%% 
This approximate formula for the $\nu_{\tau}$ mass allows 
to understand qualitatively the dominant dependence on the
SUGRA parameter space.
As we can see, the  $\nu_{\tau}$ mass is proportional to
$(1+t_{\beta}^2)$ and $\mu_3/\mu$. 
Once we have fixed the SUGRA parameters $M_0$,$A_0$ and $M_{1/2}$ 
(as is indicated in figure (\ref{fig:neut-mass})) we 
observe that the larger the value of $t_{\beta}$, the larger the
$\nu_{\tau}$ mass will be and
the smaller the value of $\mu_3/\mu$ , the smaller the 
$\nu_{\tau}$ will be, and if $\mu_3/\mu$ tends
to zero the $\nu_{\tau}$ mass tends to zero.
In figure (\ref{fig:neut-mass}) we compare for two different 
values of $t_{\beta}$ the numerical solution
(continous line), the approximate formula (\ref{eq:neumas}) 
(dashed line) (where
we make use of the definition of $s_{\zeta}$) and
our approximate formula (\ref{eq:neuapro2}) (dotted line).
We can see that the first approximation (\ref{eq:neumas}) is not perfect
and we have to sum up the second approximation which is to neglect
the contribution to $s_{\zeta}$ coming from $\delta_Bt_{\beta}$. 
After fixing the SUGRA parameters as is indicated in 
figure (\ref{fig:neut-mass}) we notice that the $\nu_{\tau}$ mass
grows with $t_{\beta}$. Apparently this is contradictory with
one of the results of ref. \cite{Nilles:1997ij} which states
that the tau neutrino decreases with increasing $t_{\beta}$.
However both results are consistent because we have included the 
implicit dependence of $\sin\zeta$ on $t_{\beta}$ while in ref.
\cite{Nilles:1997ij} $s_{\zeta}$ has been fixed.
The dependence on $t_{\beta}$ is partially cancelled by
$c^2_{\beta}$ and we obtain the $(1+t^2_{\beta})$ factor.
It was claimed in the same paper \cite{Nilles:1997ij} that 
the ratio $\mu_3/\mu$ need not be supressed in order to 
supress the tau neutrino mass. 
This afirmation depends on what one means by a 
supressed tau neutrino mass, 
because as we can see from the minimization 
equation $s_{\zeta}$ is proportional to $\mu_3/\mu$.
If we assume a tau neutrino mass of the order of the laboratory bound,
18 MeV \cite{Barate:1998,Ackerstaff:1998}, we can see in 
figure (\ref{fig:neut-mass}) that the ratio does not need 
to be supressed and it can be of the order 1.
If we assume a tau neutrino mass of the order of the cosmological bound, 
$\approx$100 eV, for a small value of $t_{\beta}$($t_{\beta}=2$) 
a ratio $\mu_3/\mu$ of order $10^{-1}$ would be allowed. 
However if we assume a tau neutrino mass of the order of 1 eV or less
(as can be infered from combinations of different experimental 
constraints \cite{Barger:1998kz}) we find for $t_{\beta}=2$ a ratio
$\mu_3/\mu$ of the order $10^{-3}$ or less.
Therefore $\mu_3/\mu$ has to be supressed 
to get neutrino masses of the eV order.
We stress that the tau neutrino mass is proportional to
the misalignement between the vectors $\mu_{\alpha}$ and $v_{\alpha}$
and this misalignement is shown through the basis invariant $s_{\zeta}$.
The unavoidable misalignement is the result of the non--universality
of the soft parameters at the electroweak scale.
Even though the non--universality is generated radiatively from the
GUT scale to the electroweak scale the parameter $s_{\zeta}$ can be of order 1
, that is to say, the misalignement can be maximum at the 
electroweak scale.
In spite of this one can get a light tau neutrino mass, from 100 eV to 
18 MeV, whithout a suppression on the ratio $\mu_3/\mu$, but if we require
a lighter neutrino the ratio $\mu_3/\mu$ needs to be suppressed.
Therefore if the tau neutrino is so light the ratio $\mu_3/\mu$
at high energy has to be suppressed too, thus giving us information about
the correct GUT model.
Finally we must point out that we can obtain a tau neutrino mass
one order of magnitud less if we set a larger value for $M_{1/2}$.\\
{\bf Calculation of $\sin\zeta$ in the three generation model}. We must add
some comments about to what extent the results would still be valid
in a three generation model. Since the trilinear terms can be very ``large''
we cannot ignore them because they contribute radiatively to the
tau neutrino mass. If we assume that there is no cancellation
between the radiative and tree level contributions to
the tau neutrino mass we can constrain the R--parity violating effects
of the bilinear terms in the same way as in the one generation model, 
i.e., through the parameter $\sin\zeta$. A crucial point 
is if the relation between the tau neutrino mass and the SUGRA 
parameter space deduced from the minimization equations
is modified. In a three generation model we can write four of the five 
minimization equations in matricial form:
 $$
M^2_L\overrightarrow{v}+
(\overrightarrow{\mu} . \overrightarrow{v})\overrightarrow{\mu}-
v_u\overrightarrow{E}+
g^2_Zv^2_Y\overrightarrow{v}=0
$$
where $\overrightarrow{\mu}$,$\overrightarrow{v}$ and 
\overrightarrow{E)} represent four vectors under 
rotation of the leptonic fields and $M^2_L$ is the $4\times4$
soft mass matrix.
Rotating to the basis II) ( where $\mu_i$ terms are zero )
we can see that we obtain a system of four linear equations for the 
sneutrino vevs in the basis II) 
and from this we can find an expression for
$\sin(\zeta)$ in a three generation model.
This relation is more complex analytically but of the
same form as the relation in the one generation model,
i.e., this relation contains terms $\Delta B_i= B_i-B_0$, terms
$\Delta M^2_i= M^2_{L_i}-M^2_{L_0}$ and the ratios $\mu_i/\mu$.
Therefore the reasoning is similar to that in the one generation
model: the parameters $\Delta B_i$ and $\Delta M^2_i$ are
not very small at the weak scale because of the perturbative bound
on $h_{top}$ thus we need to constrain the ratios $\mu_i/\mu$
if we require the tree level contribution to tau
neutrino mass be small.
%%%%%%%%%%%%%%%%%%%%%%%%%%%%%%%%%%%%%%%%%%%%%%%%%%%%%%% 
 
\section{R--parity violating effects in the scalar sector} 
 
%%%%%%%%%%%%% DISCUSION SOBRE EFECTOS DE R-PARIDAD EN EL SECTOR ESCALAR %%%% 
%%%%%%%%%%%%%%%%%%%%%%%%%% 
\begin{figure}[ht] 
\centerline{ \epsfxsize 5.8 truein \epsfbox {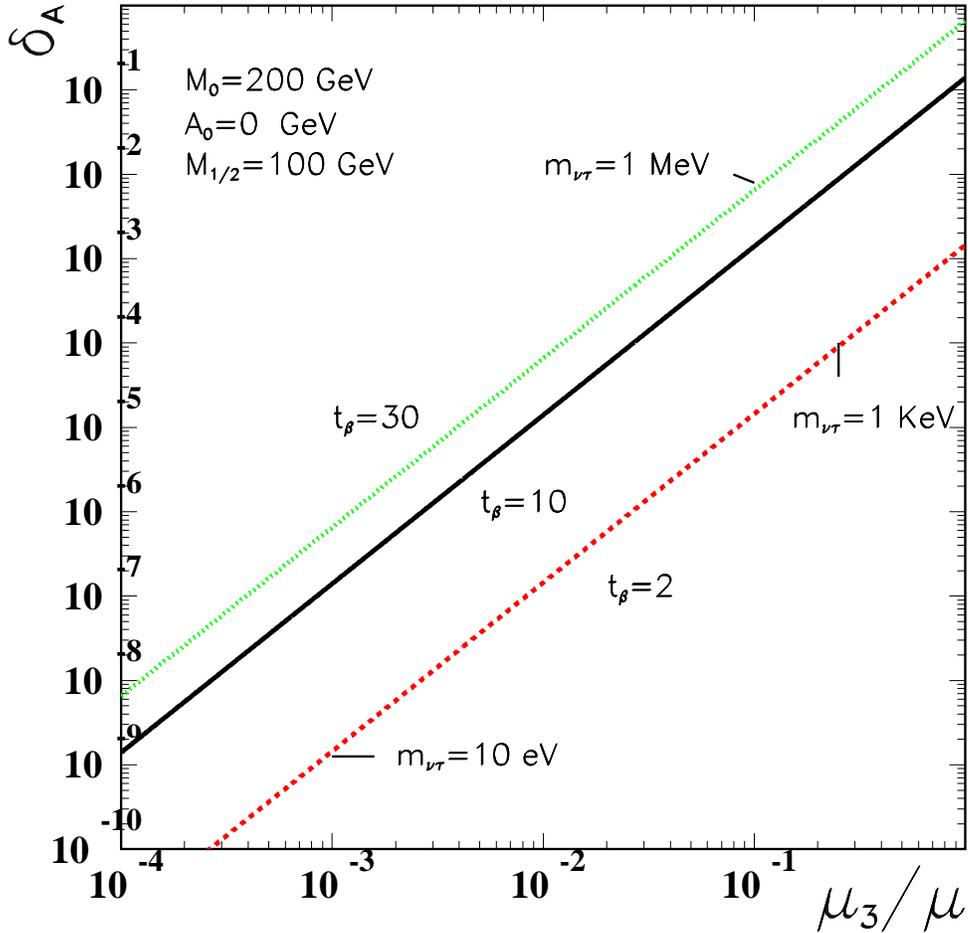}} 
\caption{\it Possible Values for $\delta_A$ as a function of the
invariant $s_{\gamma}=(\mu_3 / \mu)^I$}  
\label{fig:escalar}   
\end{figure}  
%%%%%%%%%%%%%%%%%%%%%%%%%% 
R--parity violation induces mixing effects between neutral higgses and
sneutrinos \cite{deCampos:1995av,Grossman:1997is} 
and between charged higgses and sleptons \cite{Roy:1997bu,Akeroyd:1997iq}.
We have shown in the previous section that the tau neutrino
mass is proportional to $\sin \zeta$ and  $\lambda^D_3$.
One may question if this also implies that the effects in the
scalar sector tend to zero as the tau neutrino mass tends to zero 
, or in other words, we wish to establish the relation 
between the tau neutrino mass and the R--parity
violating effects in the scalar sector.
In order to find this relation we rotate the $3\times3$ 
CP--odd neutral scalar mass matrix to the basis II) 
(where $\mu^{\prime}_3=0$) and we find:
$$ 
M_{A^0}^2= \left[  
\begin{array}{ccc} 
m_u^2 -g_Z^2v_Y^2 & \left(\frac{\mu_0^2}{\mu}B_0+\frac{\mu_3^2}{\mu} B_3 
\right) & \frac{{\mu}_0{\mu}_3}{\mu}\Delta B \\ \left(\frac{\mu_0^2}{\mu}B_0+% 
\frac{\mu_3^2}{\mu} B_3 \right) & m^{\prime 2}_{00}+g_Z^2v_Y^2 & \frac{{\mu}% 
_0{\mu}_3}{{\mu}^2}\Delta M^2 \\ \frac{{\mu}_0{\mu}_3}{\mu}\Delta B & \frac{{% 
\mu}_0{\mu}_3}{{\mu}^2}\Delta M^2 & m^{\prime 2}_{33}+g_Z^2v_Y^2  
\end{array} 
\right] \ , 
$$ 
where:  
$$ 
m^{\prime 2}_{00}+g_Z^2v_Y^2 = \frac{\mu_0^2}{\mu^2}(M_{L}^{00})^2 +\frac{% 
\mu_3^2}{\mu^2}(M_{L}^{33})^2 +\mu^2+g_Z^2v_Y^2 = (M_A^2)+\frac{\mu_3^2}{% 
\mu^2}\Delta M^2  
$$ 
\begin{equation}
m^{\prime 2}_{33}+g_Z^2v_Y^2 = \frac{\mu_0^2}{\mu^2}(M_{L}^{33})^2 +\frac{% 
\mu_3^2}{\mu^2}(M_{L}^{00})^2 +g_Z^2v_Y^2 = (M_{\widetilde{\nu}}^2)-\frac{% 
\mu_3^2}{\mu^2}\Delta M^2  
\end{equation}
Here $M_{\widetilde{\nu}}$ and $M_A$ coincide with the MSSM expressions. 
One can easily show for the CP--even mass matrix and the charged 
scalar mass matrix that the R--parity violating mixing terms are
of the form: $\mu_3\mu_0\Delta B/\mu$, $\mu_3\mu_0\Delta M^2/\mu^2$. 
In addition there are new terms proportional to $g_Z^2v^2s_{\zeta}$. 
We prefer to use the CP--odd scalar mass matrix for the sake of simplicity 
although the conclusions would be the same if we were to 
study the other scalar mass matrices.
We find the following tree level formula for the mass of the
CP-odd and for the imaginary part of the stau neutrino field 
in the R--parity violating model:
\begin{equation}
M^{\Slash{R}}_{A,\widetilde{\nu}}= \frac{1}{2} \left\{ \left( M_A^2+M_{% 
\widetilde{\nu}}^2 \right) \pm \left( M_A^2-M_{\widetilde{\nu}}^2+ 2\frac{% 
\mu_3^2}{\mu^2}\Delta M^2 \right) \left( 1+ \frac{ 2 \left(\frac{\mu_0 \mu_3% 
}{\mu^2}\Delta M^2 \right)^2} { \left( M_A^2 - M_{\widetilde{\nu}}^2 + 2% 
\frac{\mu_3^2}{\mu^2}\Delta M^2 \right) } \right)^{1/2} \right\}  
\label{eq:CP-odd-sneutrino}
\end{equation}
We define two useful parameters
$\delta_A$ and $\delta_{\widetilde{\nu}}$ as:  
$$ 
\delta_{\widetilde{\nu}}= \frac{\mu_3^2}{\mu^2}\frac{\Delta M^2} {M_{% 
\widetilde{\nu_3}}^2} \qquad \delta_A=\frac{\mu_3^2}{\mu^2}\frac{\Delta M^2}{% 
M_A^2}  
$$ 
In the limit $\delta_{A,\widetilde{\nu}}\Rightarrow 0$
we find approximate formulas which give us the mass shifts from 
their MSSM counterparts: 
\begin{equation}
(M^2_{A,\widetilde{\nu_3}})^{\Slash{R}}= (M^2_{A,\widetilde{\nu_3}}) \left\{ 
1 \pm \delta_{A,\widetilde{\nu_3}} \left( 1 + \frac{\mu_0^2}{\mu^2}\frac{% 
\Delta M^2}{(\mu^2-\Delta M^2)} \right) \right\}  
\end{equation}
Our main task is to find the relation between the parameters
$\delta_{A,\widetilde{\nu}}$ and the tau neutrino mass.
As is obvious these parameters are very similar to the soft 
parameter $\delta_M$ which appeared in the calculation of $s_{\zeta}$ 
through the minimization equations.
To get a more exact idea of the values the parameters 
$\delta_{A,\widetilde{\nu}}$ can reach we make use of 
the numerical solutions.
In figure (\ref{fig:escalar}) we show the value of $\delta_A$ 
for three different values of $t_{\beta}$, fixing the SUGRA parameters 
$M_0$, $A_0$ and $M_{1/2}$ as is indicated in the figure.
In figure (\ref{fig:escalar}) the maximum values of $\delta_A$ 
are obtained with the laboratory bound on the $\nu_{\tau}$ mass. 
As we can see, the R--parity violating effects can be large if we
assume both $t_{\beta}$ and the ratio $\mu_3/\mu$ to be large. 
Therefore large R--parity violating effects are related to neutrino masses
of the order 18 MeV.
In contrast, if we assume the minimum value for 
$t_{\beta}$ ($t_{\beta}=2$) the maximum
values for $\delta_A$ are of the order $10^{-3}$, 
even setting $\mu_3/\mu=1$. 
If we assume a neutrino mass of the order 10 eV R--parity 
violating effects in the scalar sector are in any case
negligible, $\delta_A \leq 10^{-6}$. 
 
\section{Conclusions} 
 
The phenomenology of supersymmetric R--parity violating models
has recently attracted a lot of attention
\cite{Feng:1998ad,Feng:1998qb,Diaz:1998wq,Bar-Shalom:1998xz,%
Joshipura:1998sp,Guetta:1998id,Wodecki:1998vc,Hisano:1998pe,%
Bar-Shalom:1998ap,Carena:1998gd,Carena:1998wy}.
These studies make use of different basis to 
parametrize R--parity violating effects. 
Some of the above references get bounds on sneutrino
vevs or lepton number violating $\mu$--terms while the others
put bounds on the trilinear couplings.
In this work we have shown that these bounds are misleading
unless we indicate the basis in which we are working.
We advocated, following \cite{Davidson:1998yy}, the use of 
basis independent parameters which clearly show 
the magnitude of R--parity violating effects.
As an application we have derived experimental
constraints on $\widetilde{\nu}_{\tau}$ vev from  
$Z\Rightarrow b\overline{b}$, double beta decay and 
the bounds on $\nu_e$ mass. 
We have shown that R--parity violating effects in both scalar and
fermionic sectors are controlled by the tau neutrino mass. 
By making use of the minimization equations we find a relation
between R--parity violating effects in the scalar sector and the
tau neutrino mass.
We have studied the prediction for the tau neutrino mass in a SUGRA
inspired scenario with universality of soft parameters at the GUT scale
and we have found that if we demand a very light
tau neutrino ($\leq $eV) the ratio $\mu_3/\mu$ has to be suppressed,
 although if one allows masses up to the laboratory bound
limit the ratio $\mu_3/\mu$ is unsuppresed.
We have complemented our work with a numerical study making use of
the renormalization group equations of this model and we have compared
the numerical and the analytical predictions.

\section{Acknowledgements} 
 
I would like to thank Diego Restrepo for the verification of
all the numerical results in this paper.
I would like to thank M.A.Diaz, A.Akeroyd and C.Savoy for 
beneficial discussions. 
This work was supported by DGICYT under grants PB95-1077 and by 
the TMR network grant ERBFMRXCT960090 of the European Union. 
The author is supported by a Spanish fellowship FPI of Ministerio 
de Eduaci\'on y Ciencia. 
 
%%%%%%%%%%%%%%%%%%%%%%%%%%%%%%%%%%%%%%%%%%%%%%%%%%%%%% 
 
\section{ Appendix.} 

In this appendix we give the most relevant renormalization group 
equations for this work. We make use of a condensed notation, where 
$\alpha=0,3$. 
The RGE's for trilinear couplings can be found in 
\cite{Dreiner:1995hu,Barger:1996qe} and the RGE's for
soft parameters can be found in \cite{deCarlos:1996du}.
For the trilinear interactions (Yukawas) of the superpotential 
one has:
\begin{equation} 
16 \pi^2 \frac{d h_t}{d t} = h_t \left( 6 h_t^2 + (\lambda_0^D)^2 + 
(\lambda_3^D)^2 - \left( \frac{13}{15} g_1^2 + 3 g_2^2 + \frac{% 
16}{3} g_3^2 \right) \right)  
\end{equation} 
%%%%%%%%%%%%%%%%% 
\begin{equation} 
16 \pi^2 \frac{d \lambda^D_{\alpha}}{d t} = \lambda_{\alpha}^D \left( h_t^2 
+ h_{\tau}^2 + \sum_{\alpha=0,3} 6(\lambda^D_{\alpha})^2 - 
\left( \frac{7}{15} g_1^2 + 3 g_2^2 + \frac{16}{3} g_3^2 \right) \right)  
\end{equation} 
%%%%%%%%%%%%%%%%% 
\begin{equation} 
16 \pi^2 \frac{d h_{\tau}}{d t} = h_{\tau} \left( 3(\lambda_0^D)^2 + 
4h_{\tau}^2+ 3(\lambda_3^D)^2 - \left( \frac{9}{5} g_1^2 + 3 g_2^2 \right) 
\right)  
\end{equation} 
%%%%%%%%%%%%%%%%% 

The RGE's for the soft masses relevant for our work are
(if we make use of two greek indices at the same time 
we assume they are different)
$$ 
8 \pi^2 \frac{d M_{L_{\alpha}}^2}{d t} = h_{\tau}^2 ( \sum_{\beta=0,3} 
M_{L_{\beta}}^2 + M_R^2 + A_{\tau}^2 ) + 3 (\lambda^D_{\alpha})^2 ( 
M_{L_{\alpha}}^2 + M_Q^2 + M_D^2 + A_b^2 ) +  
$$ 
\begin{equation} 
3 \lambda_{\alpha}^D \lambda_{\beta}^D M^2_{L_{\alpha\beta}} - \left( \frac{3% 
}{5}g_1^2 M_1^2 + 3 g_2^2 M_2^2 \right) - \frac{3}{10}g_1^2 {\cal S}  
\end{equation} 
%%%%%%%%%%%%%%%% 
$$ 
16 \pi^2 \frac{d M_{L_{\alpha\beta}}^2}{d t} = 
(h_{\tau}^2+3\lambda^2_{\beta}) M_{L_{\alpha\beta}}^2 + 
(3(\lambda^D_{\alpha})^2 - h_{\tau}^2 ) M_{L_{\beta\alpha}}^2 +  
$$ 
\begin{equation} 
3\lambda^D_{\alpha}\lambda_{\beta}^D (M_{L_{\alpha}}^2+M_{L_{\beta}}^2+2 
M^2_D+2M^2_Q +2A_0^DA_3^D)  
\end{equation} 
%%%%%%%%%%%%%%%%% 
Where ${\cal S}= (M_{H_u}^2 - M_{L_0}^2- M_{L_3}^2 + M_Q^2 -2 M_U^2 + M_D^2 + 
M_R^2) $. For the bilinear terms in the superpotential:
%%%%%%%%%%%%%%%%% 
\begin{equation} 
16 \pi^2 \frac{d \mu_{\alpha}}{d t} = \mu_{\alpha} \left( 3 h_t^2 +3 
(\lambda_{\alpha}^D)^2 +h_{\tau}^2 - \left( \frac{3}{5} g_1^2 + 3 g_2^2 
\right) \right) + 3\lambda^D_{\alpha}\lambda_{\beta}^D\mu_{\beta}  
\end{equation} 
%%%%%%%%%%%%%%%%% 
And for the soft bilinear terms:
%%%%%%%%%%%%%%%%% 
$$ 
16 \pi^2 \frac{d E_{\alpha}}{d t} = E_{\alpha} \left( 
3h_t^2+3(\lambda^D_{\alpha})^2+h_{\tau}^2- \frac{3}{5}g_1^2-3g_2^2 \right)+  
$$ 
\begin{equation} 
\mu_{\alpha} \left( 6h_t^2A_t+6(\lambda_{\alpha}^D)^2A_{\alpha}^D+ 
2h_{\tau}^2A_{\tau}+ \frac{6}{5}g_1^2M_1+6g_2^2M_2 \right) + 
3\lambda_{\alpha}^D \lambda_{\beta}^D \left( E_{\beta}+2\mu_{\beta} 
A_{\alpha}^D \right)  
\end{equation} 
%%%%%%%%%%%%%%%% 
As usual $g_i$ denote the gauge couplings $SU(3)_C\times 
SU(2)_L\times U(1)_Y$ (we use unification normalization: $g_1=\frac{3% 
}{5}g_Y$) and $M_i$ are gaugino mass parameters. The rest of
RGE's for: 
$A_t,A_{0}^D,A_{\tau},A_{3}^D$ and $M_{Q}^2,M_{U}^2,M^2_{D},M_{R}^2$ and $% 
M_{H_u}^2$ can be computed directly from the general formulas
\cite{deCarlos:1996du}. 
 
{\tighten

}
 
\end{document}